\documentclass[amsmath,amssymb,aip,jap,reprint]{revtex4-1}

\usepackage{graphicx}
\usepackage{dcolumn}
\usepackage{subfigure}
\usepackage{bm}
\usepackage{color}

\begin{document}

\title{On the thermal expansion in MgSiN$_{2}$}

\author{M. R{\aa}sander}\email{m.rasander@imperial.ac.uk}
\affiliation{%
Department of Materials, Imperial College London, Exhibition Road, London, SW7 2AZ, United Kingdom
}%
\author{M. A. Moram}
\affiliation{
Department of Physics, Cavendish Laboratory, University of Cambridge, JJ Thomson Avenue, Cambridge, CB3 0HE, United Kingdom
}
\date{\today}

\begin{abstract}
The thermal expansion of the wide band gap semiconductor MgSiN$_{2}$ has been determined using density functional calculations in combination with the quasi-harmonic approximation. We find that the thermal expansion is rather small in good agreement with previous experimental studies. However, the present calculations suggest that the thermal expansion of the system is more isotropic. Additional thermodynamic properties such as the Gr{\"u}neisen parameter and the heat capacity at constant pressure have also been determined and found to be in good agreement with available experiments. 
\end{abstract}

\maketitle

\section{Introduction}

Group II-IV nitride semiconductors are of growing interest for optoelectronics and high power electronic devices.\cite{Paudel2009,Punya2011,Quirk,Rasander2016,Punya2016} The group III-nitrides are often used in such application, however, the efficiency of devices composed of III-nitrides is not optimal for certain applications. For example, in ultra-violet light emitting diodes (UV-LEDs) based on III-nitrides the efficiency becomes significantly limited the shorter the wavelength of the desired UV-radiation becomes.\cite{Kneissl2011} This is due to a combination of problems in the UV-LED device design based on high content AlN semiconductor materials, such as high dislocation densities due to lattice mismatch at heterostructure interfaces and the creation of electric fields due to polarisation mismatch. The II-IV nitrides have been proposed as new types of materials in order to solve these difficulties. These materials are formed by removing the group III element in a III-nitride, e.g. AlN, and replace it with one group II, e.g. Mg, and one group IV element, e.g. Si or Ge. The bonding and crystal structures are therefore related to those of the III-nitrides but the II-IV nitrides offer different combinations of band gaps and lattice parameters and thereby opening up additional possibilities for device design. For example, Zn-based II-IV nitrides are of current interest for solar cells,\cite{Paudel2009,Punya2011} whereas wide band gap II-IV nitrides, such as MgSiN$_{2}$\cite{Quirk,Rasander2016,Punya2016} and MgGeN$_{2}$\cite{Punya2016} may find applications as part of UV optoelectronic devices.
\par
The II-IV nitrides, and MgSiN$_{2}$ in particular, form in an orthorhombic wurtzite-derived crystal structure which belongs to the Pna2$_{1}$ space group (No. 33),\cite{David1970,Bruls1} where all atoms occupy 4$a$ Wyckoff crystal positions and the primitive unit cell contains 4 formula units of MgSiN$_{2}$. The local bonding arrangement consist of distorted tetrahedrons in order to accommodate the two types of cations with different ionic radius compared to the local bonding in ideal wurtzite crystal structures.\cite{Bruls1,Rasander2017} The orthorhombic structure can be obtained by a simple transformation in the $xy$-plane of the wurtzite lattice vectors according to ${\bm a} =  {\bm a}_{1} + {\bm a}_{2}$ and ${\bm b} = 2{\bm a}_{1}$, while keeping the ${\bm c}$ lattice vector the same. Here ${\bm a}_{1}$ and ${\bm a}_{2}$ are the in-plane lattice vectors of the wurtzite crystal structure while ${\bm a}$ and ${\bm b}$ are the in-plane lattice vectors in the orthorhombic structure. The final structure of, e.g., MgSiN$_{2}$ will have $|{\bm a}|\approx\sqrt{3}|{\bm a_{1}}|$ and $|{\bm b}|\approx2|{\bm a_{1}}|$
For this orthorhombic structure, it is possible to evaluate a wurtzite-like lattice constant as $2a_{wz} = \left( a/\sqrt{3} + b/2\right)$, and it has been found that the wurtzite-like lattice constant in MgSiN$_{2}$ is only slightly larger than the lattice constant in wurtzite AlN and therefore the in-plane lattice matching between MgSiN$_{2}$ and AlN is very good with a mismatch of less than 1\%.\cite{Rasander2017}
\par
In order for materials to become useful in new technologies, a wide range of materials properties have to be investigated. For example, it is important to understand the thermal expansion such that materials in a heterostructure will have similar expansion as the temperature varies to avoid cracking. It is also important to understand the thermal conductivity of the system so that heat will flow efficiently in order to avoid overheating and eventual breakdown of semiconductor components. The thermal expansion of MgSiN2$_{2}$ has been investigated experimentally and found to be rather anisotropic as a result of the orthorhombic wurtzite-derived crystal structure of MgSiN$_{2}$.\cite{Bruls1,Bruls2} Furthermore, it was also suggested that the $a$ lattice constant has a zero or even slightly negative thermal expansion for temperatures less than 100~K. 
\par
The focus of the present study is to increase the understanding of the thermal expansion in MgSiN$_{2}$ by complementing the previous experimental studies with new density functional (DF) calculations employing the quasi-harmonic approximation (QHA) to treat the thermal expansion. In a previous study,\cite{Rasander2017} we analysed the structure and lattice dynamics of MgSiN$_{2}$. For example it was found that the phonon dispersions in MgSiN$_{2}$ were much more complex than in wurtzite AlN and that the highest frequency found for the A$_{1}^{\rm LO}$ mode at 996.4~cm$^{-1}$ is more than 100~cm$^{-1}$ higher than the highest frequency E$_{1}^{\rm LO}$ mode in AlN.\cite{Rasander2017} Furthermore, the heat capacity at constant volume and vibrational entropy were found to be in very good agreement with experiment.\cite{Rasander2017} These results were also in agreement with another theoretical study focusing on the lattice dynamics in MgSiN$_{2}$.\cite{Pramchu2017} In addition to the thermal expansion of MgSiN$_{2}$ we will here present additional physical properties available using the QHA, such as the heat capacity at constant pressure and the Gr{\"u}neisen parameter. 

\section{Theory}\label{sec:methods}
\par
The thermal expansion is evaluated within the quasi-harmonic approximation (QHA). Within the QHA, the free energy of the system is described according to
\begin{equation}\label{eq:free-energy}
F(T,V) = U(V) + E_{ZP}(V) - TS(T,V),
\end{equation}
where $U(V)$ is the internal energy of the system, as evaluated using standard DF calculations, $E_{ZP}$ is the zero point energy, evaluated as
\begin{equation}\label{eq:ZPE}
E_{ZP} = \frac{1}{N}\sum_{{\bf k},i}\hbar\omega_{{\bf k},i}(V),
\end{equation}
 $V$ is the volume and $T$ is the temperature. $\omega_{{\bf k},i}$ in Eqn.~(\ref{eq:ZPE}) is the frequency of the phonon mode $i$ at ${\bf q}$-point ${\bf k}$. $S(T,V)$ is the vibrational entropy evaluated as
\begin{multline}\label{eq:entropy}
S(T,V) = -\frac{1}{N}\sum_{{\bf k},i}k_{B} {\rm ln} \left[1- e^{\left(-\frac{\hbar\omega_{{\bf k},i}(V)}{k_{B}T}\right)}\right]\\
+ \frac{1}{N}\sum_{{\bf k},i} k_{B} \frac{\hbar\omega_{{\bf k},i}(V)}{k_{B}T}\left[ e^{\left(\frac{\hbar\omega_{{\bf k},i}(V)}{k_{B}T}\right)} -1\right]^{-1}.
\end{multline}
We note that the only term in Eqn.~(\ref{eq:free-energy}) which depends on temperature is the entropy, all remaining terms only depend on the temperature implicitly via the volume. The volume as a function of temperature, $V(T)$, is found by minimising the free energy, $F(T,V)$, with respect to the volume for any given temperature. 

\begin{table}[tb]
\caption{\label{tab:structure} Lattice constants and volume of MgSiN$_{2}$. The values labelled DFT are obtained minimising the forces at $T=0$~K and includes no zero-point energy effects. The experimental data are taken from Bruls {\it et al.}\cite{Bruls1} The experimental data given as $T=0$~K was actually measured at $T=10$~K.}
\begin{ruledtabular}
\begin{tabular}{lcccc}
XC & $a$ (\AA) & $b$ (\AA) & $c$ (\AA) & $V$ (\AA$^3$)\\
\hline
\multicolumn{5}{c}{DFT} \\
PBE & 5.3106  & 6.4950 & 5.0274 & 173.41 \\
PBEsol & 5.2768 & 6.4694 & 4.9930 & 170.45\\
\hline
\multicolumn{5}{c}{$T=0$~K} \\
PBE & 5.3303 & 6.5145 & 5.0472 & 175.26 \\
PBEsol & 5.2945 & 6.4861 & 5.0108 & 172.07\\
Expt.\cite{Bruls1} & 5.27078(5) & 6.46916(7) & 4.98401(5) &  169.9425(28) \\
\hline
\multicolumn{5}{c}{$T=300$~K} \\
PBE & 5.3333 & 6.5175 & 5.0502 & 175.54 \\
PBEsol & 5.2973 & 6.4888 & 5.0137 & 172.33\\
Expt.\cite{Bruls1} & 5.27249(4) & 6.47334(6) & 4.98622(4) & 170.1827(24)\\
\hline
\multicolumn{5}{c}{$T=500$~K} \\
PBE & 5.3392 & 6.5236 & 5.0561 & 176.11\\
PBEsol & 5.3031 & 6.4944 & 5.0195 & 172.88\\
\end{tabular}
\end{ruledtabular}
\end{table}
\par
The volume thermal expansion, $\alpha_{V}$, is defined as
\begin{equation}
\alpha_{V} = \frac{1}{V}\frac{dV}{dT}.
\end{equation}
Since MgSiN$_{2}$ has an orthorhombic structure where the volume is given by $V= abc$, where $a$, $b$ and $c$ are the lattice constants along the $x$, $y$ and $z$ crystal directions, the volume thermal expansion can be expressed as
\begin{equation}\label{eq:ind-expansion}
\alpha_{V} = \frac{1}{a}\frac{da}{dT} + \frac{1}{b}\frac{db}{dT} + \frac{1}{c}\frac{dc}{dT} = \alpha_{a} + \alpha_{b} + \alpha_{c},
\end{equation}
where $\alpha_{X}$ is the thermal expansion of the $X$ axis of the crystal, with $X$ being $a$, $b$ or $c$. By assuming an isotropic material it is possible to define a linear thermal expansion $\alpha_{L}$ by
\begin{equation}\label{eq:linear-expansion}
\alpha_{L} = \frac{1}{3}\left(\alpha_{a} + \alpha_{b} + \alpha_{c}\right) = \frac{\alpha_{V}}{3}.
\end{equation}
We note that the above definition of the linear thermal expansion, $\alpha_{L}$, could still be useful in the case of non-cubic materials. For such anisotropic solids, $\alpha_{L}$ measures the average thermal expansion in any of the $x$, $y$ and $z$ crystal directions.
\par
\section{Computational details}\label{sec:comp}
\par
Density functional (DF) calculations have been performed using the projector augmented wave (PAW) method\cite{Blochl} as implemented in the Vienna {\it ab initio} simulations package (VASP).\cite{KresseandFurth,KresseandJoubert} We have used the two generalized gradient approximations PBE\cite{PBE} and PBEsol\cite{PBEsol} for the exchange-correlation energy functional. The plane wave energy cut-off was set to 800~eV and we have used $\Gamma$-centered k-point meshes with the smallest allowed spacing between k-points of 0.1~\AA$^{-1}$. The atomic positions and simulation cell shapes were relaxed until the Hellmann-Feynman forces acting on atoms were smaller than 0.0001~eV/\AA.
\par
The QHA calculations were performed using a grid of volumes around the equilibrium volume obtained using the DF calculations, with an increased density of points close to the equilibrium volume, see Fig.~\ref{fig:e-v}. For each volume the shape of the unit cell, i.e. the relative length of the lattice vectors, and the positions of the atoms are fully relaxed as described above. Phonon calculations were performed for this volume grid using the Phonopy code\cite{Chaput2011,Parlinski1997} with supercells based on a 2$\times$2$\times$2 repetition of the 16 atom MgSiN$_{2}$ primitive unit cell. We note that there is essentially no difference in the phonon dispersions obtained using this size of the supercell compared to supercells based on a 3$\times$3$\times$3 repetition of the MgSiN$_{2}$ unit cell. For further details regarding the evaluation of the lattice dynamics see Ref.~\onlinecite{Rasander2017}.
\par
To evaluate the thermal expansion along the $x$, $y$ and $z$ crystal directions it is required to first obtain the individual lattice constants as a function of temperature for each direction. The temperature dependence of the lattice constants are evaluated from the temperature dependence of the volume, $V(T)$, which is evaluated within the QHA on the previously mentioned volume grid. The volumes at specific temperatures, e.g. at $T=10$~K, 100~K and 300~K, are then used as input to obtain the lattice constants $a$, $b$ and $c$ at these specified temperatures. This is done by relaxing the geometry of the lattice vectors and internal atomic coordinates as mentioned previously for these fixed volumes. The lattice constants are then fitted to the expression $X(T) = x_{0} + x_{2}T^2 + x_{3}T^3$, where $X$ is either $a$, $b$ or $c$, and the thermal expansion coefficient along $X$, $\alpha_{X}$, is calculated using Eqn.~(\ref{eq:ind-expansion}). In this fitting procedure, we have restricted the temperatures to be within 0 and 500~K.

\section{Results}
\begin{figure}[tb]
\includegraphics[width=8cm]{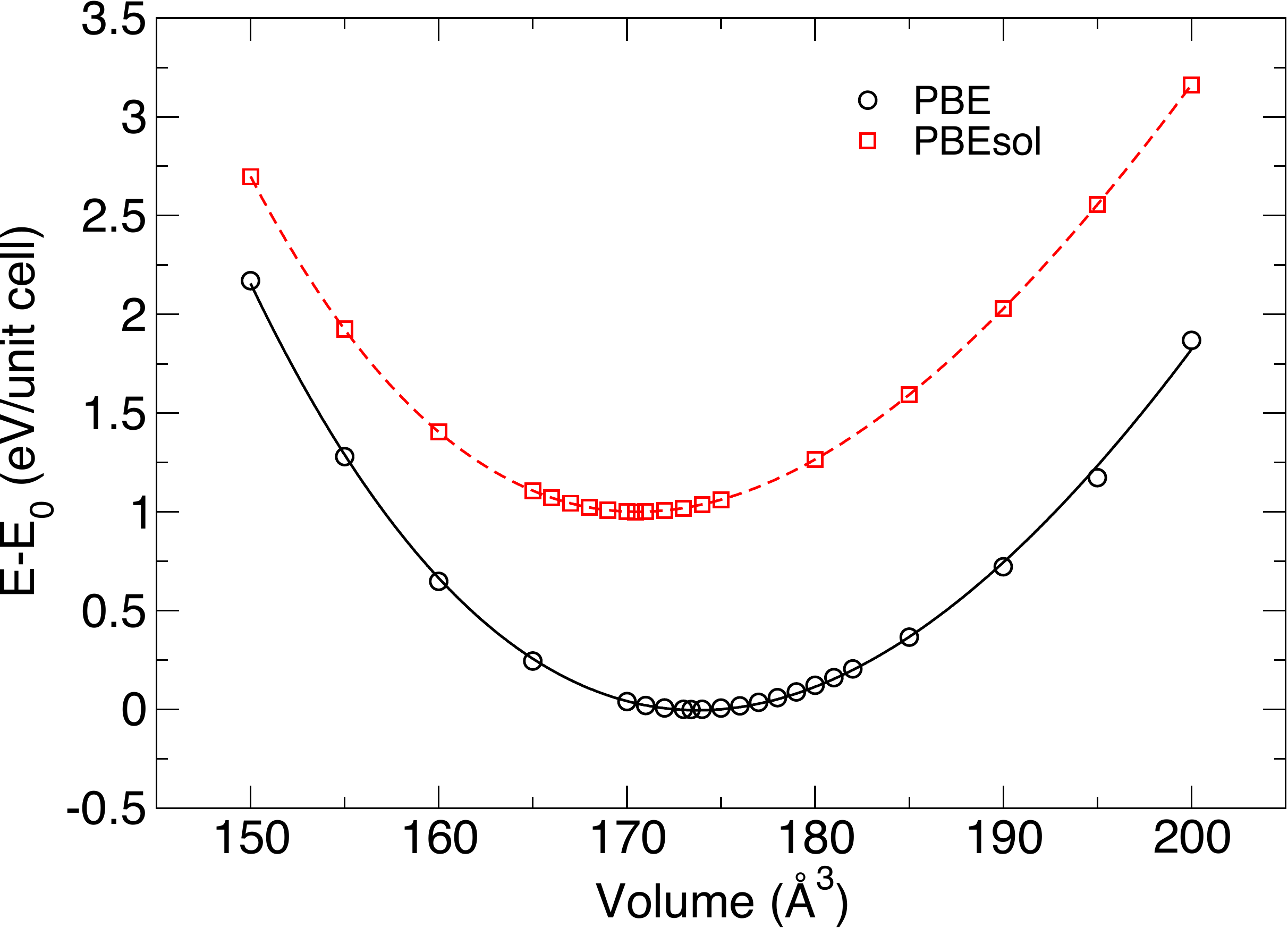}
\caption{\label{fig:e-v} (Color online) Calculated energy versus volume of MgSiN$_{2}$ using the PBE and PBEsol approximations. The energies obtained using the PBEsol approximations have been shifted by +1~eV.}
\end{figure}
In Fig~\ref{fig:e-v} we show the calculated energy as a function of volume using both PBE and PBEsol approximations. The obtained equilibrium volumes are 173.41~\AA$^3$ and 170.45~\AA$^3$ for the PBE and PBEsol approximations, respectively. Table~\ref{tab:structure} shows the calculated and experimental volumes and lattice constants at $T=0$ and 300~K. We note that the equilibrium volume and lattice constants derived from a standard DF calculation do not contain any vibrational effects. When vibrational effects are taken into consideration, for example through the use of the QHA, the $T=0$~K volume is increased compared to the standard DFT result due to zero point phonon effects (ZPPE).\cite{Csonka2009,Rasander2015} 
Here we include the effect of the zero-point energy (ZPE) of Eqn.~(\ref{eq:ZPE}) which increases the size of the equilibrium structure compared to pure DF calculations as shown in Table~\ref{tab:structure}. Experimental low temperature structure analysis has obtained the volume to be 169.94~\AA$^3$ at $T=10$~K. This implies that both the standard DFT and the QHA $T=0$~K results are larger than the experimental value for both PBE and PBEsol approximations. The same holds at $T=300$~K where the experimental structure is slightly smaller than the theoretical results. We note that the PBEsol approximation is closer to experiment than the PBE approximation, which is to be expected, and that the deviation between theory and experiment is as expected for this type of semiconductor material.\cite{Rasander2015}
\par
We note that there are two experimental studies of the thermal expansion of MgSiN$_{2}$ in the literature: The first study by Bruls {\it et al.}\cite{Bruls1} provide the thermal expansion along $a$, $b$ and $c$ directions as well as the linear thermal expansion evaluated according to Eqn.~(\ref{eq:linear-expansion}). In this study the thermal expansion was obtained by measuring the variation of the lattice constants using time-of-flight neutron powder diffraction between 10 and 300~K. The second study,\cite{Bruls2} also by Bruls {\it et al.}, provides the linear thermal expansion between 300~K and 1600~K, where the thermal expansion was measured using a dual rod dilatometer in nitrogen using sapphire as the reference material. Due to experimental constrains the second study only provide the linear thermal expansion.\cite{Bruls2} We note that the linear thermal expansion obtained in these two studies differ slightly, e.g. at 300~K the first study obtained $\alpha_{L}=4.4$~K$^{-1}$\cite{Bruls1} while the second study obtained $\alpha_{L}=3.82$~K$^{-1}$.\cite{Bruls2}
\par

\begin{figure*}[t]
\includegraphics[width=15cm]{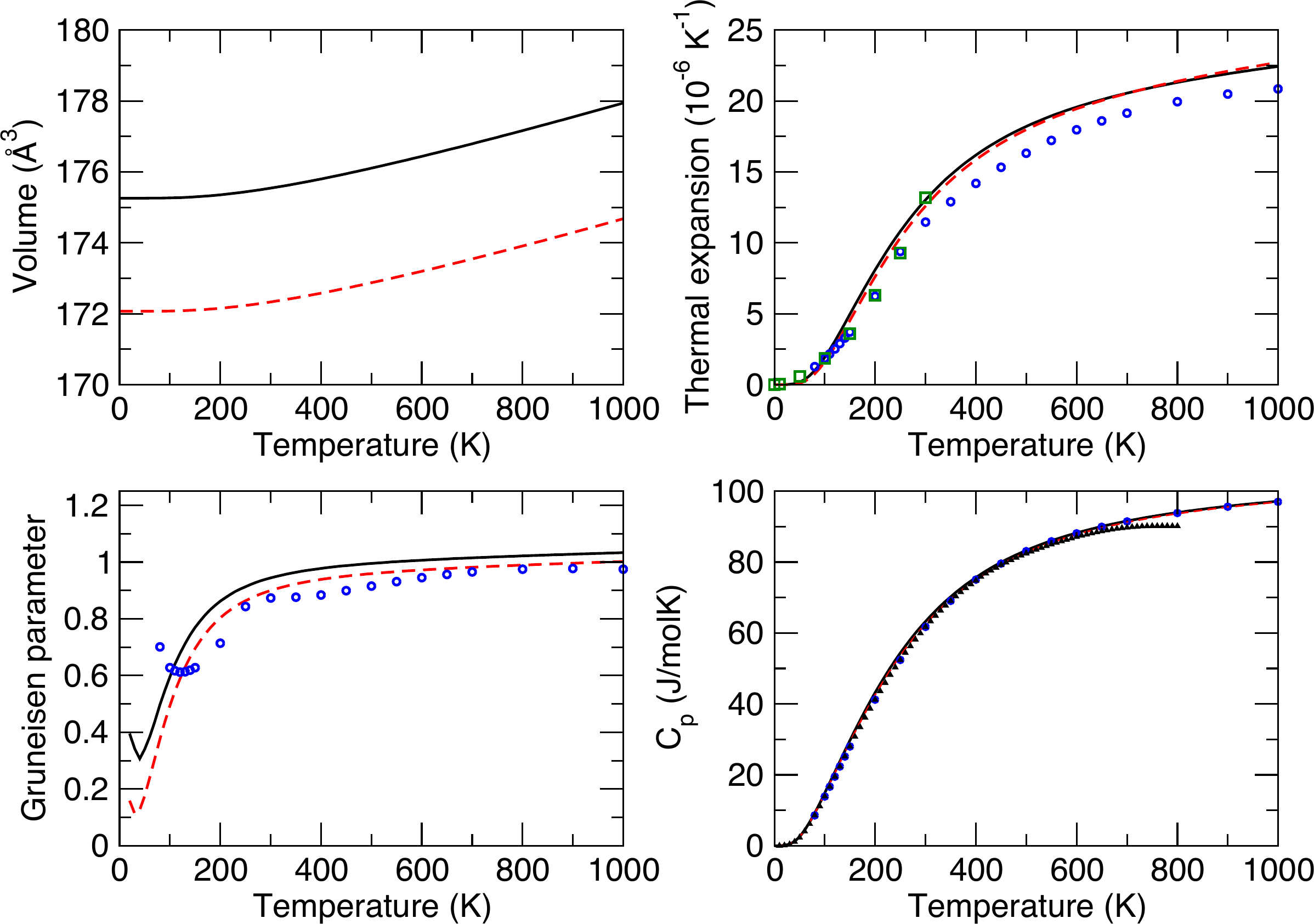}
\caption{\label{fig:expansion-plot} (Color online) Calculated volume, thermal expansion, Gr{\"u}neisen parameter and heat capacity at constant pressure of MgSiN$_{2}$ as a function of temperature obtained using the PBE (solid) and PBEsol (dashed) approximations. Experimental data taken from Bruls {\it et al.}\cite{Bruls1} (green squares), Bruls {\it et al.}\cite{Bruls2} (blue circles) and Bruls {\it et al.}\cite{Bruls1998} (black triangles). Note: The thermal expansion shown in the upper right panel is the volume thermal expansion $\alpha_{V}$.}
\end{figure*}
\begin{figure}[t]
\includegraphics[width=8cm]{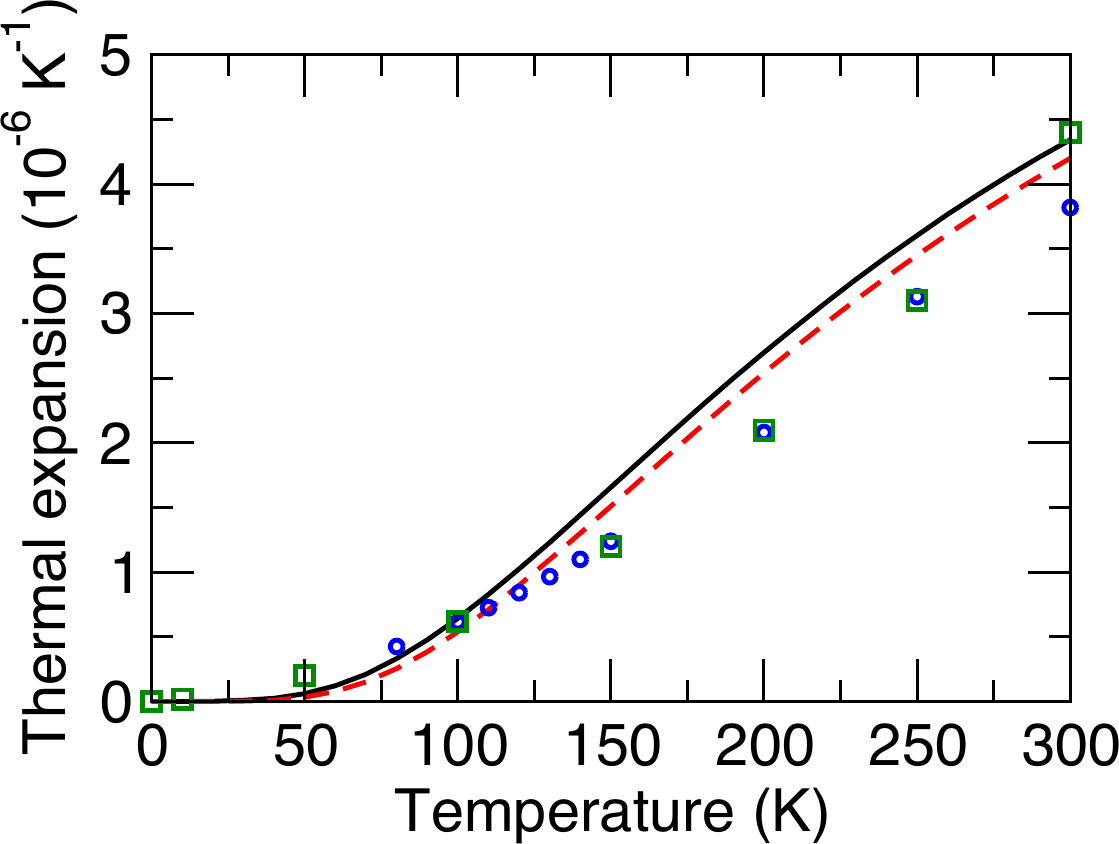}
\caption{\label{fig:alpha} (Color online) Experimental and calculated linear thermal expansion up to 300~K. The PBE and PBEsol results are shown using solid (black) and dashed (red) lines. Experimental data from Bruls {\it et al.} are shown using (green) squares\cite{Bruls1} and (blue) circles.\cite{Bruls2}}
\end{figure}
Fig.~\ref{fig:expansion-plot} shows the calculated volume, thermal expansion, Gr{\"u}neisen parameter and heat capacity at constant pressure as functions of temperature between 0 and 1000~K along with experimental data for the thermal expansion,\cite{Bruls1,Bruls2} Gr{\"u}neisen parameter\cite{Bruls2} and the heat capacity.\cite{Bruls2,Bruls1998} We note that the volume expansion is very small at low temperatures with hardly any change in the crystal volume below 200~K. Even though the PBE and PBEsol approximations results in different volumes the variation with temperature is very similar for both approximations. Furthermore, we find the volume thermal expansion to be in good agreement with experimental thermal expansion data, even though the calculated volumes are larger compared to the experimental values. We also find the calculated Gr{\"u}neisen parameter to be in good agreement with experiment, especially the PBEsol data at higher temperatures are in great agreement with experiment.  In addition, the calculated heat capacity data are in excellent agreement with available experimental data.
\par
Fig.~\ref{fig:alpha} shows the linear thermal expansion coefficient from both theory and experiment up to $T=300$~K. Here it is made clear that the calculated linear expansion coefficients are smaller than the experimental expansion coefficient at low temperatures, below 100~K, and larger compared to experiment for temperatures above 100~K.
\par
\begin{figure}[t]
\includegraphics[width=8cm]{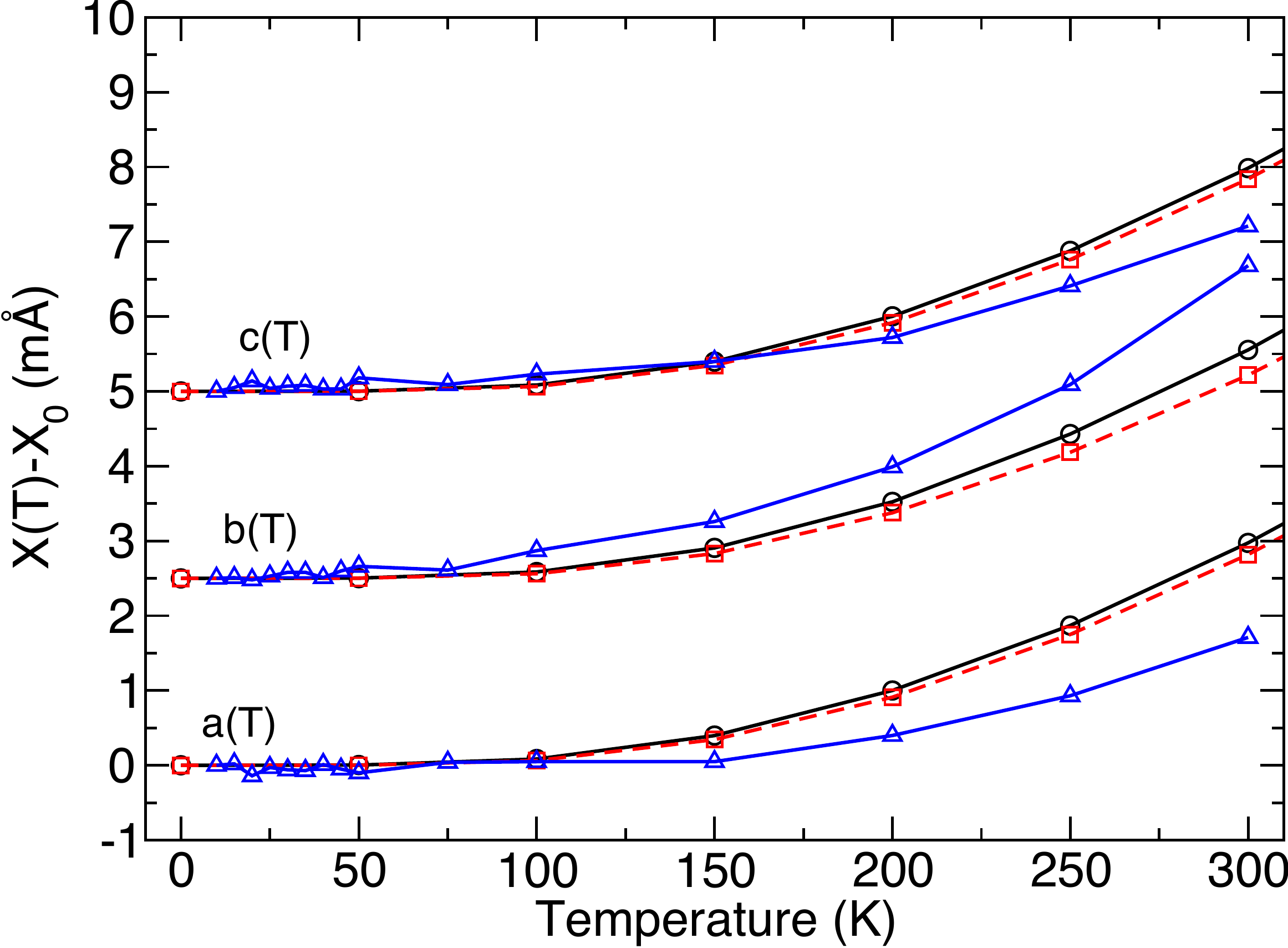}
\caption{\label{fig:lat-const} (Color online) Expansion of the lattice constants, $a$, $b$ and $c$ as a function of temperature in the range between $T=0$~K and $T=300$~K. The lattice constants are in reference to corresponding value at $T=0$~K. The $b(T)$ and $c(T)$ lattice constants have been shifted by 2.5 and 5~m\AA, respectively. Experimental values are taken from Bruls {\it et al.}\cite{Bruls1}}
\end{figure}

In Fig.~\ref{fig:lat-const} we show the expansion of the lattice constants as a function of temperature between 0 and 300~K. The theoretical lattice constants are obtained from the corresponding volumes at specific temperatures obtained using the QHA as discussed in Section~\ref{sec:comp}. The calculated expansion of the lattice constants are in overall good agreement with available experiments. We note that PBE and PBEsol approximations provide rather similar expansions, which is also the case for the calculated volume thermal expansion. We find that theory and experiment agrees very well in the case of the out-of-plane lattice constant $c$. In addition, we  find that the calculations result in a larger expansion along the $a$ axis compared to experiment and a smaller expansion along the $b$ axis compared to experiment. Overall, the calculations provide a more isotropic expansion of the lattice constants compared to the experimental result of Bruls {\it et al.}\cite{Bruls1} which is more anisotropic. We also note that, even though the expansion is very small or even negligible for low temperatures, the calculations do not support a reduction of the $a$ lattice constant at low temperatures as indicated by Bruls {\it et al.}\cite{Bruls1}
\par
To provide a more detailed understanding of the thermal expansion, we provide in Table~\ref{tab:alpha} the calculated and experimental linear thermal expansion coefficients as well as the thermal expansion coefficients along the $a$, $b$ and $c$ crystal directions. As can be seen in Table~\ref{tab:alpha} the calculated thermal expansion along $a$ and $c$ are consistently overestimated compared to experiment, however, the agreement becomes better as the temperature increases. The thermal expansion along the $b$ direction is in very good agreement with experiment at lower temperatures, but as the temperature increases the thermal expansion along $b$ is consistently smaller than the experimental results. Once again the calculated results suggest a more isotropic expansion of the crystal.
\begin{table*}[tb]
\caption{\label{tab:alpha} Calculated thermal expansion along $a$, $b$ and $c$ directions as well as the average linear thermal expansion compared to experiment. Note: Two different values are shown for the calculated linear thermal expansion as described in the text. Also note: The experimental thermal expansions along $a$, $b$ and $c$ directions and linear thermal expansion evaluated through $\frac{1}{3}\left(\alpha_{a}+\alpha_{b}+\alpha_{c} \right)$ are taken from Bruls {\it et al.}\cite{Bruls1} while the experimental linear thermal expansion data presented as $\alpha_{L}$ are taken from Bruls {\it et al.}\cite{Bruls2}}
\begin{ruledtabular}
\begin{tabular}{lcccccc}
XC & $\alpha_{a}$  & $\alpha_{b}$  & $\alpha_{c}$ & $\alpha_{wz}$ &  $\frac{1}{3}\left(\alpha_{a}+\alpha_{b}+\alpha_{c} \right)$ & $\alpha_{L}$ \\
 & ($\times$10$^{-6}$ K$^{-1}$) & ($\times$10$^{-6}$ K$^{-1}$) & ($\times$10$^{-6}$ K$^{-1}$) & ($\times$10$^{-6}$ K$^{-1}$)  & ($\times$10$^{-6}$ K$^{-1}$) & ($\times$10$^{-6}$ K$^{-1}$)\\
\hline
\multicolumn{7}{c}{$T=10$~K} \\
PBE & 0.12 & 0.10 & 0.13 & 0.11 & 0.12 & 0.00 \\
PBEsol & 0.11 & 0.09 & 0.12 & 0.10 & 0.11 &  0.00 \\
Expt. & -0.02\cite{Bruls1} & 0.06\cite{Bruls1} & 0.03\cite{Bruls1} & 0.02 & 0.02\cite{Bruls1} & - \\
\hline
\multicolumn{7}{c}{$T=100$~K} \\
PBE & 1.26 & 1.06 & 1.34 & 1.16 & 1.22 & 0.64 \\
PBEsol & 1.18 & 0.93 & 1.26 & 1.06 &  1.12 &  0.54\\
Expt. & 0.25\cite{Bruls1} & 1.0\cite{Bruls1} & 0.60\cite{Bruls1} & 0.64 & 0.62\cite{Bruls1} & - \\
\hline
\multicolumn{7}{c}{$T=200$~K} \\
PBE & 2.63 & 2.21 & 2.78 & 2.41 & 2.54 & 2.70 \\
PBEsol & 2.52 & 1.98 & 2.68 & 2.24 &  2.39 & 2.54\\
Expt. & 1.5\cite{Bruls1} & 2.8\cite{Bruls1} & 1.9\cite{Bruls1} & 2.17 & 2.1\cite{Bruls1} & - \\
\hline
\multicolumn{7}{c}{$T=300$~K} \\
PBE & 4.09 & 3.43 & 4.33 & 3.75 & 3.95 &  4.34\\
PBEsol & 3.99 & 3.14 & 4.25 & 3.55 &  3.79 & 4.20\\
Expt. & 3.7\cite{Bruls1} & 5.5\cite{Bruls1}  & 4.0\cite{Bruls1} & 4.63 & 4.4\cite{Bruls1}  & 3.82\cite{Bruls2} \\
\end{tabular}
\end{ruledtabular}
\end{table*}
\par
Since the linear thermal expansion can be determined via the volume thermal expansion, which is rigorously determined in the QHA framework, as well as via the sum of the expansions along the crystal directions, see Eqn.~(\ref{eq:linear-expansion}), it makes sense to provide and to compare the results using both these approaches. We find that the calculated linear thermal expansion evaluated as the sum of the thermal expansions in the three crystal directions is significantly higher than linear expansion obtained using the volume expansion for low temperatures. As the temperature increases the two values become more similar. Note that the linear thermal expansion evaluated directly from the volume thermal expansion through the QHA calculations is the one which is more accurate. The difference between the two approaches is likely due to the very small variations in volume taking place for low temperatures. Small volume variations with temperature implies small variations in the energy and forces during relaxations such that the difference in lattice constants as the temperature is varied is small. For each temperature, small variations of the lattice constants will be degenerate and lead to small inaccuracies in determining the lattice constants as a function of temperature for small temperatures. Errors present in the method to determine the individual lattice constants as a function of temperature will therefore be more significant for these small temperatures where small variations of the structure occur. These errors will also affect the quality of the curve fitting performed to obtain the thermal expansion coefficients along the crystal directions. Overall, however, we conclude that the calculated thermal expansion coefficients are in good agreement with experimental results.
\par
Since the MgSiN$_{2}$ structure is derived from the wurtzite crystal structure it is possible to obtain a wurtzite-like lattice constant, $a_{wz}$, in the $xy$-plane of the MgSiN$_{2}$ structure as
\begin{equation}\label{eq:wurtzite-lat}
a_{wz} = \frac{1}{2}\left[ \frac{a}{\sqrt{3}} + \frac{b}{2}\right].
\end{equation}
This is the lattice constant that is the most useful when comparing the MgSiN$_{2}$ structure with materials that form in the wurtzite crystal structures, such as AlN.\cite{Rasander2017} Based on Eqn.~(\ref{eq:wurtzite-lat}) it is possible to derive an expression for the thermal expansion of the wurtzite-like lattice constant in terms of the thermal expansion of the $a$ and $b$ lattice constants:
\begin{equation}\label{eq:alpha-wz}
\alpha_{wz} = \frac{1}{2a_{wz}} \left[ \frac{a}{\sqrt{3}}\alpha_{a} + \frac{b}{2}\alpha_{b} \right].
\end{equation}
The thermal expansion coefficients of the wurtzite-like lattice constant are shown in Table~\ref{tab:alpha}.
\par
It is now possible to compare the thermal expansion coefficients of MgSiN$_{2}$ directly with the coefficients for crystals in the wurtzite structure, such as AlN and GaN. For AlN, Yim and Paff\cite{Yim1974} have found the average thermal expansion coefficient in the range 20 to 800$^{\circ}$C along the $c$ axis to be 4.2$\times10^{-6}$/K and the corresponding in-plane expansion coefficient to be 5.3$\times10^{-6}$/K. Figge {\it et al.}\cite{Figge2009} have found the expansion coefficients to be slightly larger with high temperature limits for the expansion coefficients of 5.8$\times10^{-6}$/K and 7.1$\times10^{-6}$/K, for the $a$ and $c$ lattice constants in AlN, respectively. Common to both these studies is that the expansion coefficient for the in-plane lattice constant $a$ is larger than the coefficient for the out-of-plane lattice constant $c$. According to the experimental results for MgSiN$_{2}$ obtained by Bruls {\it et al.}\cite{Bruls1} it seems that the expansion coefficient of the in-plane wurtzite-like lattice constant $a_{wz}$ is larger than the expansion coefficient of the out-of-plane lattice constant, as shown in Table~\ref{tab:alpha}, which is a similar behaviour to what is observed in AlN. The expansion coefficients derived from the calculations, as shown in Table~\ref{tab:alpha}, suggest that the out-of-plane expansion coefficient is larger than the expansion coefficient for the wurtzite-like lattice constant. This is due to the overestimation of the calculated expansion coefficient along the $c$-axis compared to the measured values as discussed previously. Overall, when comparing MgSiN$_{2}$ with AlN it is found that the thermal expansion coefficients in MgSiN$_{2}$ are larger than in AlN. This is clearly shown by Bruls {\it et al.}\cite{Bruls2} for the average linear expansion coefficient and our calculations support this fact. 
\begin{figure}[t]
\includegraphics[width=8cm]{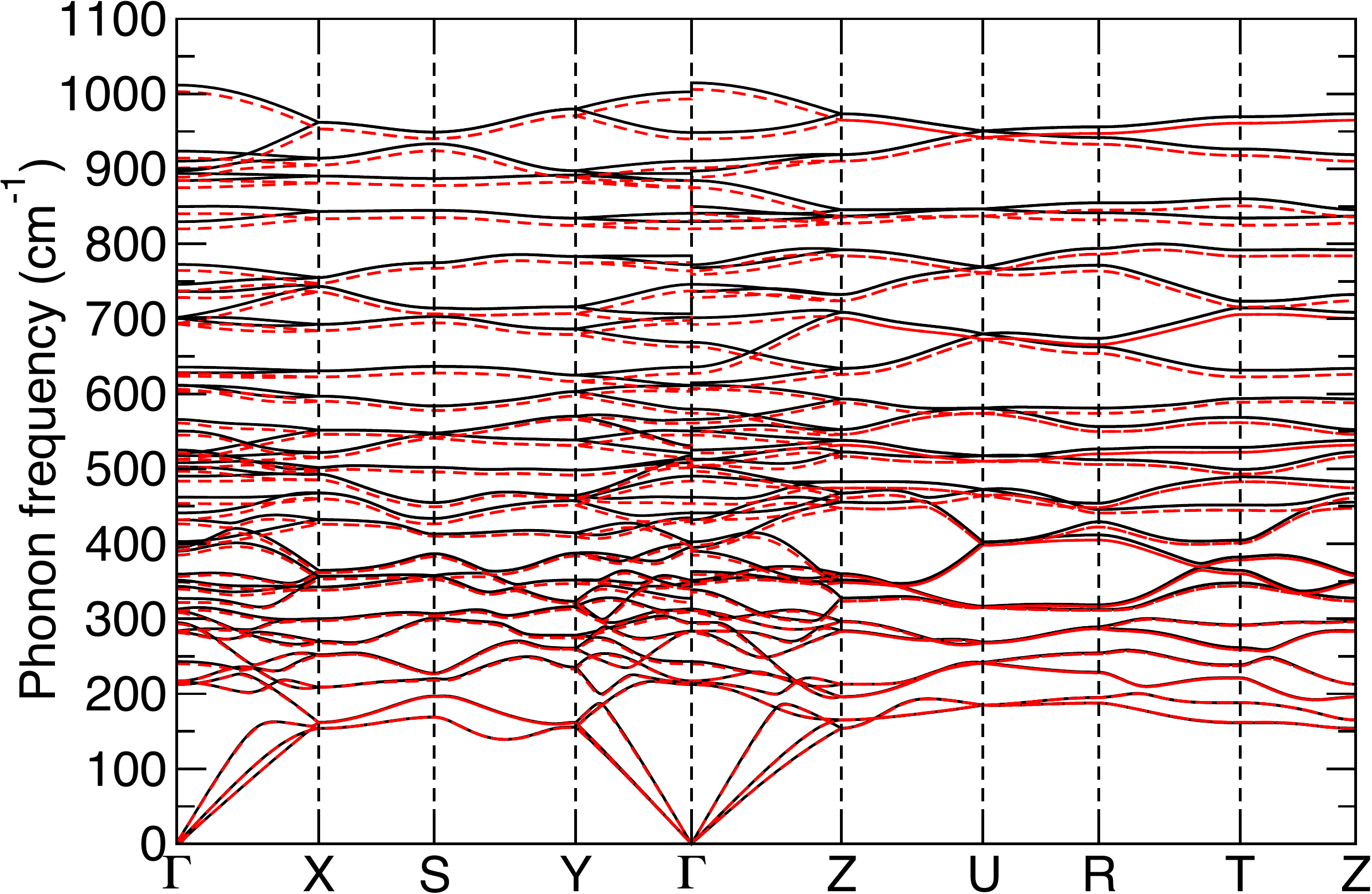}
\caption{\label{fig:modes} (Color online) Calculated phonon modes for the volumes $V=170.45$~\AA$^3$ (solid black) and 172.07~\AA$^3$ (dashed red).}
\end{figure}
\par
In Fig.~\ref{fig:modes} we show the calculated phonon modes along high symmetry directions in the Brillouin zone as obtained using the PBEsol approximation. The phonon dispersions in MgSiN$_{2}$ have been discussed in detail by R{\aa}sander {\it et al.}\cite{Rasander2017} and will only be briefly discussed here. These phonon dispersions are in good agreement with already published phonon dispersions based on the PBE approximation.\cite{Rasander2017} The difference is that the PBEsol approximation yields slightly higher frequencies compared to the PBE approximation. The highest frequency found for the A$_{1}^{\rm LO}$ using PBEsol is 1014.4~cm$^{-1}$ compared to 996.4~cm$^{-1}$ as obtained using the PBE approximation. This is a results of the improvement in bonding obtained when using the PBEsol approximation for solids in comparison to using the PBE approximation.\cite{Rasander2015} The highest frequency found using the PBEsol approximation is also in great agreement with the highest frequency found using Raman spectroscopy of a MgSiN$_{2}$ powder, where the highest frequency was found to be 1026~cm$^{-1}$.\cite{Rasander2017}
\par
In Fig.~\ref{fig:modes} there are phonon modes obtained using two different volumes: The first is the equilibrium volume as obtained from a pure DF calculation, i.e. including no ZPPE. The second is the $T=0$~K volume obtained by including ZPPE in the calculation through the QHA. The increased volume softens the modes, especially so for higher frequency modes. When the volume is increased as a result of increased temperature, the softening of the modes will continue. However, the change is very small when considering the modes obtained for volumes corresponding to $T=0$~K and $T=300$~K. In fact, the change is much smaller between $T=0$~K and $T=300$~K compared to the change shown in Fig.~\ref{fig:modes}. Hence, there are no dramatic changes in the lattice dynamics as the temperature is increased, which leads us to conclude that the lattice expansion is driven by a continuous variation of the free energy of the system as the temperature is increased.  
\begin{figure}[t]
\includegraphics[width=8cm]{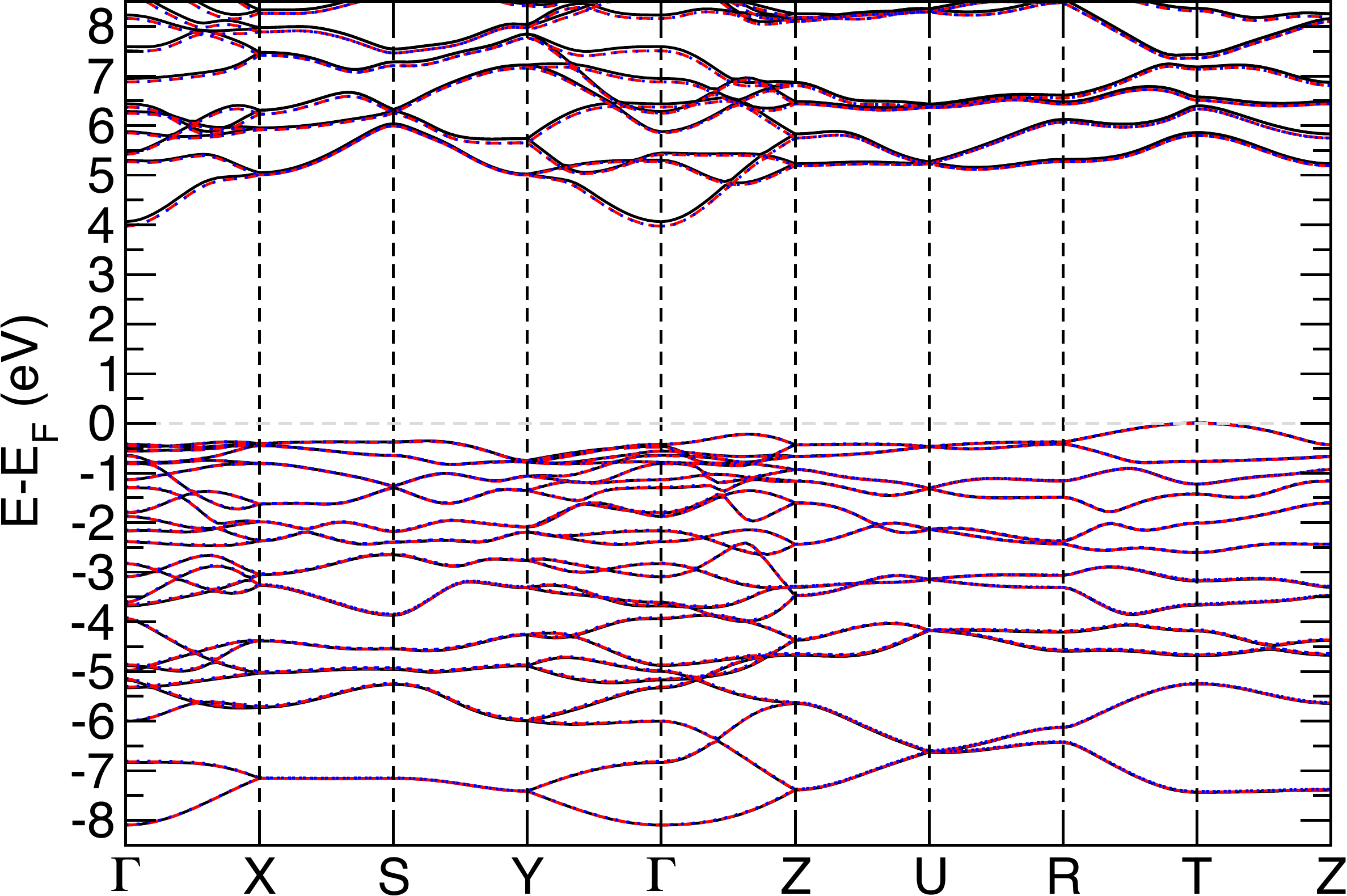}
\caption{\label{fig:bands} (Color online) Calculated energy energy bands for the volumes $V=170.45$ (solid black), 172.07 (dashed red) and 172.33~\AA$^3$ (dotted blue).}
\end{figure}
\par
It is also interesting to see if there are any variations in the band structure of the system with increasing temperature. Therefore, Fig.~\ref{fig:bands} shows the calculated band structure obtained using the PBEsol approximation for three different volumes corresponding to the equilibrium DFT volume and volumes corresponding to $T=0$ and $T=300$~K. Note that we only discuss changes to the electronic structure obtained with the increase of the temperature within the QHA. For further details of the electronic structure of MgSiN$_{2}$ see Refs.~\onlinecite{Quirk,Punya2016}. As the temperature increases the change is even less dramatic than the variation of the phonon modes. There is essentially no change in the features of the valence bands, and for the lowest conduction bands we only find a very small lowering of the conduction band close to the conduction band minima at the $\Gamma$-point. Therefore it is clear that there are no significant changes in the band structure with temperature when the temperature effects are simulated using the QHA. 

\section{Summary and conclusions}
We have calculated the thermal expansion and other thermodynamic properties of MgSiN$_{2}$. It is found that the thermal expansion is small, especially at low temperatures. and comparable to the thermal expansion in wurtzite AlN. The calculated thermal expansion is also found to be in good agreement with previous experimental studies. The main difference between the calculated thermal expansion presented here and the previous measured expansion is that the calculated thermal expansion is more isotropic compared to the experimental results which are more anisotropic. We also find no significant changes in the lattice dynamics as the temperature increases; neither do we find any significant changes in the band structure as an effect of increasing temperature. 

\section{Acknowledgements}
We acknowledge support from the Leverhulme Trust via M. A. Moram's Research Leadership Award (RL-007-2012). M. A. Moram acknowledges further support from the Royal Society through a University Research Fellowship. This work used the Imperial College high performance computing facilities and, via our membership of the UK's HEC Materials Chemistry Consortium funded by EPSRC (EP/L000202), the ARCHER UK National Supercomputing Service (http://www.archer.ac.uk).

\bibliography{mgsin2}

\end{document}